\documentclass[aip,cha,amsmath,reprint]{revtex4-1}
\usepackage{graphicx}
\usepackage{amssymb}
\usepackage[absolute]{textpos}
\usepackage{amsfonts}
\usepackage{xspace}

\newcommand{\rhoc}[1]{\ensuremath{\rho_{#1}^\mathrm{c}}\xspace}
\newcommand{\rhom}[1]{\ensuremath{\rho_{#1}^\mathrm{m}}\xspace}

\newcommand {\ac} {\ensuremath{a_\mathrm{c}}\xspace}
\newcommand {\am} {\ensuremath{a_\mathrm{m}}\xspace}
\newcommand {\hac}[1] {\ensuremath{\hat{a}_\mathrm{c}(#1)}\xspace}
\newcommand {\ham}[1] {\ensuremath{\hat{a}_\mathrm{m}(#1)}\xspace}

\newcommand {\acr} {$a^{(r)}_\mathrm{c}$\xspace}
\newcommand {\amr} {$a^{(r)}_\mathrm{m}$\xspace}
\newcommand {\bacd} {$\bar{a}^{(D)}_\mathrm{c}$\xspace}
\newcommand {\bamd} {$\bar{a}^{(D)}_\mathrm{m}$\xspace}
\newcommand {\bac} {$\bar{a}_\mathrm{c}$\xspace}
\newcommand {\bam} {$\bar{a}_\mathrm{m}$\xspace}

\newcommand {\eacr} {\ensuremath{\widetilde{a}^{(r)}_\mathrm{c}}\xspace}
\newcommand {\eamr} {\ensuremath{\widetilde{a}^{(r)}_\mathrm{m}}\xspace}
\newcommand {\bacr} {\ensuremath{\overline{\widetilde{a}}^{(r)}_\mathrm{c}}\xspace}
\newcommand {\bamr} {\ensuremath{\overline{\widetilde{a}}^{(r)}_\mathrm{m}}\xspace}

\newcommand{\ER}{Erd\H{o}s-R\'{e}nyi\xspace}
\newcommand{\adj}[1]{\ensuremath{A_{#1}}}
\newcommand{\Adj}{\ensuremath{\boldsymbol{A}}\xspace}

\begin{document}

\title{Assortative mixing in functional brain networks during epileptic seizures}

\author{Stephan Bialonski}
\email{bialonski@gmx.net}
\affiliation{Max Planck Institute for the Physics of Complex Systems, N\"othnitzer Stra{\ss}e~38, 01187~Dresden, Germany}
\author{Klaus Lehnertz}
\email{klaus.lehnertz@ukb.uni-bonn.de}
\affiliation{Department of Epileptology, University of Bonn, Sigmund-Freud-Stra{\ss}e~25, 53105~Bonn, Germany}
\affiliation{Helmholtz Institute for Radiation and Nuclear Physics, University of Bonn, Nussallee~14--16, 53115~Bonn, Germany}
\affiliation {Interdisciplinary Center for Complex Systems, University of Bonn, Br{\"u}hler Stra{\ss}e~7, 53175~Bonn, Germany}

\received{28 June 2013}
\accepted{8 September 2013}
\published{20 September 2013}

\begin{abstract}
We investigate assortativity of functional brain networks before, during, and after one-hundred epileptic seizures with different anatomical onset locations.
We construct binary functional networks from multi-channel electroencephalographic data recorded from 60 epilepsy patients; and from time-resolved estimates of the assortativity coefficient we conclude that positive degree-degree correlations are inherent to seizure dynamics.
While seizures evolve, an increasing assortativity indicates a segregation of the underlying functional network into groups of brain regions that are only sparsely interconnected, if at all. Interestingly, assortativity decreases already prior to seizure end.
Together with previous observations of characteristic temporal evolutions of global statistical properties and synchronizability of epileptic brain networks, our findings may help to gain deeper insights into the complicated dynamics underlying generation, propagation, and termination of seizures.
\end{abstract}
\maketitle

\begin{textblock*}{15cm}(3cm,26cm)
Copyright (2013) American Institute of Physics. This article may be downloaded for personal use only. Any other use requires prior permission of the author and the American Institute of Physics. The article appeared in Chaos 23, 033139 (2013) and may be found at \url{http://link.aip.org/link/doi/10.1063/1.4821915} --- DOI: 10.1063/1.4821915
\end{textblock*}

\begin{quotation}
Epilepsy is a disorder of the brain characterized by an enduring predisposition to generate epileptic seizures. 
It affects more than 50 million individuals worldwide, and for 25\,\% of epilepsy patients, seizures remain poorly controlled despite maximal medical management.
There is increasing evidence that focal-onset seizures---that appear to originate from a circumscribed region of the brain---result from complex interactions in a large-scale epileptic network, which comprises cortical and subcortical brain structures and regions. 
Specific temporal changes of characteristics of epileptic brain networks were hypothesized to reflect an emergent self-regulatory mechanism for seizure termination. 	
We aimed at shedding more light onto this mechanism as it would provide important clues as to how to efficiently control networks underlying seizure dynamics.
To do so, we investigate---in a time-resolved manner---the assortativity of epileptic networks, i.e., the preference for brain regions (nodes) to interact with other brain regions with similar properties.  
Epileptic networks show assortative mixing patterns. These patterns change in a characteristic way and indicate that assortativity increases during seizures, reaches a maximum prior to the end of seizures, and decreases again at the end of seizures. We speculate that these changes may reflect a reorganization of the underlying functional brain networks, which may result in seizure termination.
\end{quotation}

\section{Introduction}

Complex networks\cite{Newman2003,Boccaletti2006a,Arenas2008,BarratBook2008} have been recognized to be powerful representations of complex systems and can advance our understanding of their dynamics. 
They are studied in diverse disciplines ranging from earth\cite{Abe2006,Abe2006b,Jimenez2008} and climate science\cite{Tsonis2004,Donges2009,Steinhaeuser2011} to the neurosciences\cite{Reijneveld2007,Bullmore2009,Sporns2011a,Stam2012b}. 
In this perspective, systems are considered to be composed of subsystems (i.e., nodes) which can or cannot interact with each other according to some underlying physical coupling topology (represented by links between nodes). 
While such structural networks have been studied extensively and are considered to serve as the physical substrate on which dynamical patterns can emerge, it is only recently that research into network dynamics gained strong momentum\cite{Arenas2008,BarratBook2008}. 
The system dynamics may be represented by an interaction (or functional) network in which nodes represent subsystems and links reflect interactions between them. 
In field studies, such networks are usually derived via time series analysis techniques where nodes are associated with sensors and links are derived from the strength and/or direction of interactions as assessed by estimators of signal interdependence\cite{Pikovsky_Book2001,Pereda2005,Hlavackova2007,Lehnertz2009b,Lehnertz2011b}.

Using concepts from network theory, a plethora of methods have been developed to probe and assess diverse global and local properties of networks such as clustering coefficient, average shortest path length, synchronizability, or the degree distribution. 
Moreover, numerous studies demonstrated that network topologies can decisively depend on the tendency of links to connect nodes with certain local properties\cite{Boccaletti2006a,BarratBook2008}. 
For instance, if people of a group tend to make friends with people speaking the same language, then this group is likely to separate into distinct circles of friends according to languages\cite{Newman2003b}. 
This tendency of links to connect nodes with similar or equal properties is called \emph{assortativity} and was frequently studied with respect to node degrees (i.e., the number of links connected to nodes). 
Assortativity according to node degrees can be quantified by the assortativity coefficient\cite{Newman2002a,Newman2003b} and is considered to reflect the extent of degree-degree correlations present in the network topology. 
If links preferentially connect nodes of similar (dissimilar) degree, such networks are called assortative (disassortative). 
The presence of degree-degree correlations was found to have far-reaching consequences for network resilience (disassortative networks are more vulnerable to attacks than assortative networks)\cite{Newman2002a,Newman2003b}, for the ability of a network to globally synchronize (disassortative networks appear to be easier to synchronize than assortative ones)\cite{Bernardo2005,Motter2005a,Newth2005}, and the tendency of a network to separate into distinct groups (assortative networks show a stronger tendency to disintegrate into different groups than disassortative networks)\cite{Newman2003b}. 
Moreover, in assortative networks, hubs (nodes with relatively high degree) tend to be closely interconnected thereby forming a resilient core which may facilitate the spread of information over the network.

Many technological and biological networks appear to be disassortative\cite{Newman2003c}, while social\cite{Newman2003c}, seismic\cite{Abe2006b}, and functional brain networks derived from measurements of neural activity (e.g., electroencephalogram, magnetoencephalogram, or functional magnetic resonance imaging data)\cite{Eguiluz2005,Park2008,Wang2010c,Kramer2011,Jalili2011,Moussa2011,Braun2012} were reported to be assortative. 
Structural brain networks may exhibit assortativity\cite{Hagmann2008,Bassett2008,Bassett2011a,Moussa2011} or disassortativity\cite{Park2008} depending on the investigated scale\cite{Bettencourt2007,vandenHeuvel2011},  
and other factors had recently been identified that might affect assortativity\cite{Jalili2011,Pernice2011,Schwarz2011,Joudaki2012,Anderson2013}. 
Studies that investigated the impact of neurological or neurodegenerative diseases on structural or functional brain networks reported on increased assortativity in patients suffering from Alzheimer's disease\cite{deHaan2009}, dementia\cite{Agosta2013} or schizophrenia\cite{Bassett2008}, in patients with brain tumors\cite{Wang2010c}, with autism spectrum disorder\cite{Tsiaras2011} or with psychogenic non-epileptic seizures\cite{Barzegaran2012} as compared to healthy controls.

Epilepsy represents one of the most common neurological disorders, affecting approximately 1\,\% of the world's population\cite{Duncan2006}. In about 25\,\% of individuals with epilepsy, seizures cannot be controlled by any available therapy.
Early conceptualizations of epileptic seizures as either focal or generalized\cite{Engel1989} are being challenged by an increasing evidence of seizure dynamics (generation, spread, and termination) within a network of brain regions (so called epileptic network)\cite{Spencer2002,Lehnertz2009b,Lemieux2011,Berg2011,Lehnertz2013} which generate and sustain normal, physiological brain dynamics during the seizure-free interval. 
Using concepts from network theory, previous studies reported on a specific temporal evolution of global properties (clustering coefficient, average shortest path length, synchronizability) of epileptic networks\cite{Ponten2007,Kramer2008,Schindler2008a,Ponten2009,Kramer2010,Kuhnert2010,Bialonski2011b} during seizures. 	
This finding is of importance for improving our understanding of seizure dynamics in humans, given the similarity of topological evolution across different types of epilepsies, seizures, medication, age, gender, and other clinical features.

In order to provide a more complete view, here we report on the temporal evolution of assortativity of functional brain networks underlying epileptic seizures.
We derive these networks from multichannel electroencephalograms which were recorded prior to, during, and after $100$ epileptic seizures from 60 patients (cf. section~\ref{sec:methods}). 
We assess the robustness of our findings by comparing results obtained from networks derived using two different estimators of signal interdependence. 
Moreover, we demonstrate in a simulation study (cf. Appendix~A) that degree-degree correlations can be spuriously induced by the finite size and frequency content of time series and may not necessarily reflect properties of the underlying dynamics. 
Thus, we compared our results with those obtained from random network ensembles specifically designed to account for the aforementioned influences. 
After reporting results in section~\ref{sec:results}, we conclude with interpreting our findings and how they may relate to results obtained in earlier studies (section~\ref{sec:conclusion}).

\section{Data and methods}
\label{sec:methods}
We retrospectively analyzed electroencephalographic recordings of focal onset seizures from 60 patients with drug-resistant epilepsy. 
Seizure onsets were localized in different anatomical regions. 
All patients had signed informed consent that their clinical data might be used and published for research purposes. 
The study protocol had been approved by the ethics committee of the University of Bonn. 
Electroencephalograms (EEGs) were recorded prior to, during, and after 100 epileptic seizures as reported in previous studies \cite{Schindler2008a,Bialonski2011b}. 
The multichannel ($53\pm21$ channels) EEGs were recorded via chronically implanted strip, grid or depth electrodes from the cortex and/or from within relevant brain structures.
Data were sampled at 200\,Hz, digitized using a 16-bit analog-to-digital converter, and filtered within a frequency range of 0.5--70\,Hz. 
A bipolar referencing was applied prior to subsequent steps of analysis\cite{Schindler2008a}. 
Electroencephalographic seizure onset and seizure end was automatically determined\cite{Schindler2007a}.

We pursued a sliding window approach which allowed for a time-resolved analysis of the evolution of assortativity of seizure-related functional brain networks.
We divided each multichannel recording into non-overlapping consecutive windows of length $T=500$ sampling points ($2.5$\,s duration), and for each channel we normalized data from each window to zero mean and unit variance. 
In order to derive a functional network for each window, we associated each channel $i$ with a node $i$ and derived links by estimating signal interdependencies between all pairs of time series.
We employed two commonly utilized methods to assess signal interdependence. 
The first method makes use of the absolute value of the correlation coefficient between time series at nodes $i$ and $j$:
\begin{equation}
 \rho_{ij}^\mathrm{c} = \left| T^{-1} \sum_{t=1}^{T} (x_i(t)-\bar{x}_i)(x_j(t)-\bar{x}_j)\hat{\sigma}_i^{-1}\hat{\sigma}_j^{-1} \right| \mbox{,} 
\label{eq:cc}
\end{equation}
where $x_i$ denotes the time series at node $i$ with mean $\bar{x}_i$ and estimated standard deviation $\hat{\sigma}_i$.

The second method can take into account possible time lags (e.g., due to propagation of electrical activity along anatomical pathways during the seizure) when assessing signal interdependencies and is defined as the maximum value of the absolute cross correlation function,
\begin{equation}
 \rho_{ij}^\mathrm{m} = \max_{\tau} \left\{ \left| \frac{\xi(x_i,x_j)(\tau)}{\sqrt{\xi(x_i,x_i)(0)\xi(x_j,x_j)(0)}} \right| \right\},
\label{eq:mc}
\end{equation}
with
\begin{equation}
\xi(x_i,x_j)(\tau) = \begin{cases} \sum_{t=1}^{T-\tau} x_i(t+\tau) x_j(t)  &, \tau \geq 0 \\ \xi(x_j,x_i)(-\tau) &, \tau < 0\text{,} 
\end{cases}
\end{equation}
where $\tau$ denotes the time lag. 
Both estimators are symmetric ($\rhoc{ij}=\rhoc{ji}$ and $\rhom{ij}=\rhom{ji}$), are confined to the interval $[0,1]$, and  assess linear dependencies of time series. Here we refrain from assessing nonlinear dependencies since previous studies\cite{Horstmann2010,Hartman2011} reported results obtained from estimators of nonlinear and of linear signal dependence to be qualitatively similar when pursuing network analyses of neuroscientific data.

Links can now be defined via thresholding the values of signal interdependence such that the adjacency matrix \Adj of the functional network has entries
\begin{equation}
\adj{ij} = \begin{cases}
 1, & \rho_{ij}>\theta, i\neq j\\
 0, & \text{otherwise,}
\end{cases}
\end{equation}
where $\theta\in[0,1]$ denotes the threshold and $\rho_{ij}$ either denotes \rhoc{ij} or \rhom{ij}. 
We choose $\theta$ such that the resulting network possesses a predefined link density\cite{Anderson1999}
\begin{equation}
 \epsilon = \frac{\bar{k}}{(N-1)}\text{,}
\end{equation}
with the number of nodes $N$ and the mean degree $\bar{k} = N^{-1}\sum_{i=1}^N k_i$ ($k_i$ denotes the degree of node $i$). In the following, we set $\epsilon = 0.1$ to define links.

To assess assortativity of the functional networks we employ the assortativity coefficient $a$, which is defined as the Pearson correlation coefficient between the degrees of nodes at both ends of a link\cite{Newman2002a,Newman2003b}. Reformulating the Pearson correlation coefficient in terms of the degrees of 
nodes\cite{Lehnertz2013} (cf. Appendix B), we obtain
\begin{equation}
 a = \left(K_1 K_3 - K_2^2 \right)^{-1} \left(2K_1 \sum_{i=1}^N \sum_{j=1}^{i-1} \adj{ij}k_i k_j -K_2^2 \right) \text{,}
\label{eq:assort}
\end{equation}
where $K_u = \sum_{i=1}^N k_i^u$, and $a$ is confined to the interval $[-1,1]$ by definition. 
Positive (negative) values of $a$ indicate an assortative (disassortative) network. 
Note that the assortativity coefficient is not well defined for regular graphs, i.e., networks whose nodes share all the same degree.
In the following, we denote the assortativity coefficient of the functional networks with $a_c$ or $a_m$ depending on whether signal interdependencies are estimated by the correlation coefficient (Eq.~\ref{eq:cc}) or by the maximum value of the absolute cross correlation function (Eq.~\ref{eq:mc}).

It was demonstrated in previous studies\cite{Bialonski2011b,Palus2011} that the finite length of empirical time series and their frequency content together with the applied analysis methodology to infer networks can introduce spurious properties in interaction networks. 
These properties are not related to the analyzed dynamics but reflect the way how networks are derived from finite multichannel data. 
Since the dynamics of epileptic seizures is well known for its complex temporal changes in frequency content\cite{Franaszczuk1998b,Schiff2000,Jouny2003,Bartolomei2010}, it is important to investigate whether and to which extent the assortativity coefficient reflects spurious properties induced by the analysis methodology. 
We detail the findings of our investigation in the appendix (cf. Appendix~A) and briefly note here that the assortativity coefficient takes on larger positive values the larger the relative amount of low-frequency contributions and the shorter the length of time series. 

In order to account for these influences and to distinguish them from those reflecting the dynamics, we compare the assortativity coefficient of a functional network with those obtained from an ensemble of random networks\cite{Bialonski2011b}. 
To generate a random network, we proceed as above but estimate signal interdependencies between pairs of surrogate time series\cite{Schreiber2000a,Bialonski2011b}, which preserve the length of the EEG time series from which they are derived, their frequency contents, and their amplitude distributions but are in all other aspects random. 
The surrogate time series comply with the null hypothesis of independent linear stochastic processes.
The resulting random network possesses the same link density as the functional network. 
We denote the assortativity coefficient of a random network with $a_c^{(r)}$ or $a_m^{(r)}$ depending on the applied estimator for signal interdependencies. 

\section{Results}
\label{sec:results}

\begin{figure*} 
\begin{center}
 \includegraphics[width=0.8\textwidth]{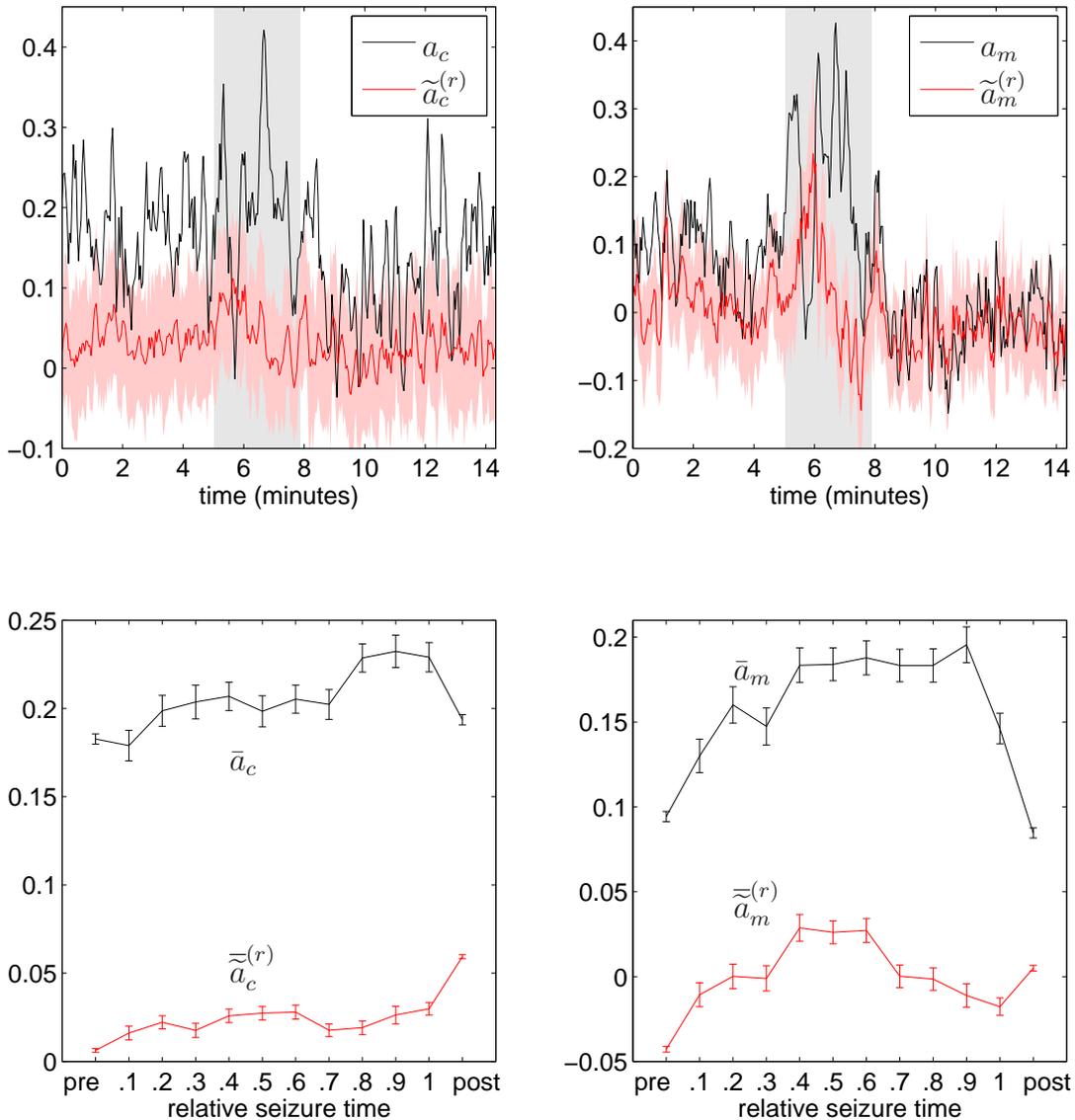}
\end{center}
\caption{Top row: Assortativity coefficients \ac and \am (black lines) during an exemplary seizure. Mean values and $\pm 1$~standard deviations of network properties obtained from surrogate time series (\eacr, \eamr) are shown as red line and red-shaded areas, respectively. Profiles are smoothed using a four-point moving average. The gray-shaded area marks the seizure. Bottom row: Mean values (black) of network properties \bac (left), \bam (right) averaged separately for pre-seizure, discretized seizure, and post-seizure time periods of 100 epileptic seizures. Mean values \bacr and \bamr are shown in red and were obtained from  ensembles of random networks. All error bars indicate standard errors of the mean. Lines are for eye-guidance only.}
\label{fig:1}
\end{figure*}

In the top panels of Fig.~\ref{fig:1}, we show time courses of the assortativity coefficients \ac (top left) and \am (top right) for an exemplary recording of a seizure. 
Both coefficients indicate that the majority of functional brain networks obtained for this recording is assortative. 
We observe \am to increase at the beginning of the seizure and to decrease at the end of the seizure. 
A similar course in time is also noticeable for \ac (which exhibits a larger variability during the recording). 
For this recording, \ac and \am indicate functional brain networks to possess a more assortative structure before than after the seizure.

In order to investigate whether these time courses could be explained by the finite size of time series and their frequency content, which varies over time, we generate---for each functional brain network---an ensemble of 20 random networks and determine the assortativity coefficients \acr and \amr. 
With \eacr and \eamr, we denote the respective mean values calculated from the ensembles. 
After the seizure and to a lesser extent also before the seizure, \eamr approximates the values of \am quite well (cf. top panels of Fig.~\ref{fig:1}). 
During the seizure, however, pronounced differences between \am and \eamr as well as between \ac and \eacr indicate that the increased assortativity cannot be related to simple alterations in frequency content nor be related to the finite size of time series but reflect features of the seizure dynamics.

To summarize our findings for all recordings of 100 focal onset seizures, we need to account for the different durations of seizures (mean seizure duration: $110\pm 60\,s$). 
To this end, we partition each seizure into ten equidistant time bins\cite{Schindler2008a,Bialonski2011b}, assign the estimated assortativity coefficients to their respective time bins, and determine, for each time bin, the mean values (\bac, \bam, \bacr, \bamr). In addition, we also determined mean values from an equal number of pre-seizure and post-seizure windows.

Despite the fact that anatomical locations of seizure onset varied across patients, a common time course of the assortativity coefficients becomes apparent (bottom panels of Fig.~\ref{fig:1}). 
\bac increases at the beginning of seizures, keeps increasing until it reaches a plateau at the end of the seizures, and decreases again after the seizures.  
A similar behavior can be observed for \bam, which increases at the beginning of the seizures, reaches a plateau already in the middle of the seizures from where it decreases again already prior to the end of seizures. 

\begin{figure}
\begin{center}
 \includegraphics[width=0.4\textwidth]{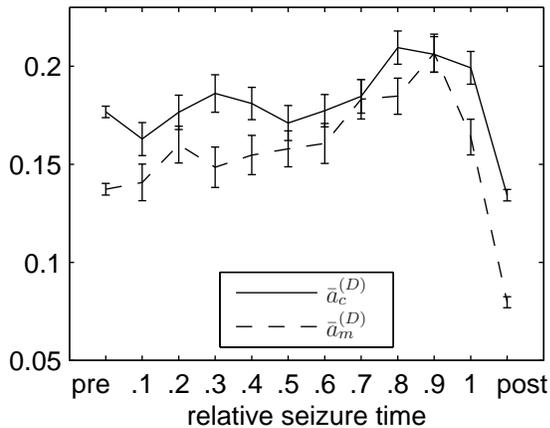}
\end{center}
\caption{Difference values $\bar{a}^{(D)}_\mathrm{c}$ and $\bar{a}^{(D)}_\mathrm{m}$ for pre-seizure, discretized seizure, and post-seizure time periods of 100 epileptic seizures. All error bars indicate standard error of the mean. Lines are for eye-guidance only.}
\label{fig:2}
\end{figure}

Time courses of \bacr and \bamr (bottom panels of Fig.~\ref{fig:1}) obtained from random networks indicate that values of \bac and \bam cannot be solely explained by the varying frequency content and the finite length of time series. 
We compare values of the assortativity coefficients with those obtained from random networks by determining the differences $\ac - \eacr$ and $\am - \eamr$ for each window. We assign these values to their respective time bins, and determine, for each time bin, their mean values ($\bar{a}_c^{(D)}$, $\bar{a}_m^{(D)}$). The time courses of \bacd and \bamd (shown in Fig.~\ref{fig:2}) are remarkably similar: both indicate an ongoing increase of assortativity during the seizures, peak in the last fifth part of the seizures, and decrease again after the seizures.

\section{Conclusion}
\label{sec:conclusion}

We pursued a time-resolved analysis of functional brain networks derived from multichannel EEG data from epilepsy patients recorded before, during, and after seizures. 
We assessed the assortativity coefficient which quantifies the tendency of nodes to preferentially connect to nodes with a similar degree (degree-degree correlations). 
Our results suggests that seizure dynamics are characterized by assortative functional networks, i.e., by positive degree-degree correlations. 
We demonstrated in a simulation study that assortative network structures can also spuriously arise due to the finite size and frequency content of time series and thus---like the clustering coefficient and the average shortest path length\cite{Bialonski2011b}---do not necessarily reflect properties of the underlying dynamics. 
By comparing with suitable random networks, however, we corrected for the aforementioned influences and still observed functional brain networks to be assortative during seizures. 
This result was obtained for two different estimators of signal interdependence that we employed to derive functional brain networks. 
Thus we conclude that positive degree-degree correlations are inherent to seizure dynamics as assessed by our analysis methodology.

We observed the assortativity coefficient to change in a characteristic way during seizures which indicates a reorganization of the underlying functional brain networks.
While functional brain networks before and after the seizures were less assortative,  assortativity slowly increased during seizures and reached a maximum prior to the end of seizures. 
Since this change was observable irrespective of the anatomical onset location, this functional reorganization might be a generic feature of focal seizure dynamics. 

\begin{figure*}
\begin{center}
   \includegraphics[width=\textwidth]{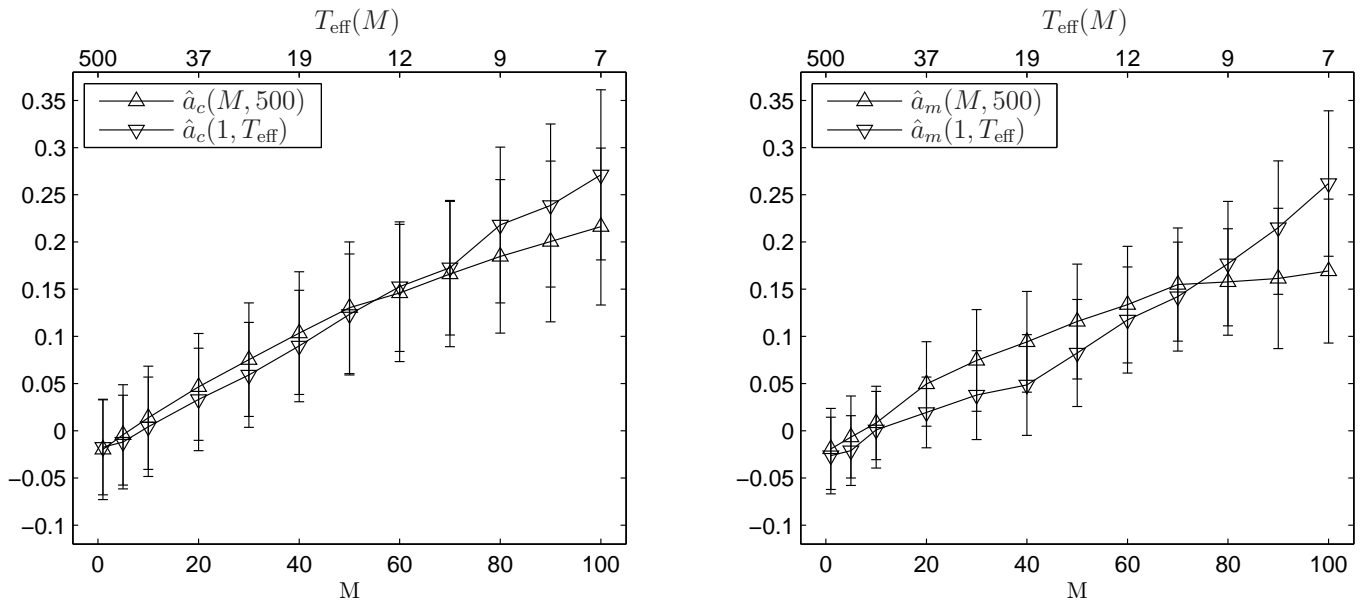}
\end{center}
\caption{Left panel: Dependence of the assortativity coefficient $\hat{a}_c(M,500)$  and $\hat{a}_c(1,T_\text{eff}(M))$ on the size $M$ of the moving average and of the length $T_\text{eff}$ of time series. Error bars indicate standard deviations obtained from ensembles of $E=1000$ networks. Right panel: Same as left panel but for the assortativity coefficients $\hat{a}_m(M,500)$  and $\hat{a}_m(1,T_\text{eff}(M))$. Lines are for eye-guidance only.}
\label{fig:3}
\end{figure*}

Our observations agree with and complement earlier findings\cite{Schindler2008a,Bialonski2011b} for the clustering coefficient and the average shortest path length of functional brain networks which were found to increase during seizures while synchronizability decreased. 
A re-increased synchronizability observed at the end of seizures might indeed promote neuronal synchronization which in turn was hypothesized to be a potential mechanism for seizure termination\cite{Schindler2007a,Schindler2007b}. 
Our study sheds light on how the topology of functional brain networks changes during seizures. These changes either promote or weaken the ability of networks to show synchronization: with increasing assortativity during seizures, the hubs of the functional brain network become mutually interconnected while the low-degree nodes become connected to each other.
This segregation of the network is likely to result in different groups of brain regions which are---if at all---only sparsely interconnected. 
The lower interconnectivity between groups is reflected in larger values of the average shortest path length, while larger values of the clustering coefficient indicate strong connectivity within groups. 
At the same time, the segregation coming along with increasing assortativity appears to weaken the synchronizability and possibly to hinder the synchronization of the network as a whole. 
This is in agreement with various numerical studies which reported synchronizability to decrease with increasing assortativity or with increasing clustering coefficient\cite{McGraw2005,Bernardo2005,Motter2005a,Newth2005,Wu2006a,Wang2007}. 
In this perspective, seizure termination might be characterized by a reintegration of groups of brain regions within the network thereby diminishing segregation which in turn may lower the threshold for the network to show global synchronization. 

Our findings can inform the development of model studies which may be able to further improve our understanding of mechanisms leading to or aborting global network synchronization. 
In addition, our study underlines the importance of hubs and their connectivity within the epileptic network. 
Identifying hubs and their role within epileptic networks could help to further our understanding of the generation, propagation, and termination of seizures and might guide the improvement of seizure prevention strategies. 

\begin{acknowledgments}
We thank G. Ansmann and F. Ghanbarnejad for helpful comments. This work was supported by the Deutsche Forschungsgemeinschaft (Grant No. LE660/4-2).
\end{acknowledgments}

\appendix
\section{Influencing factors and null models}

In a previous study\cite{Bialonski2011b}, it was demonstrated that the finite length and the frequency content of time series can introduce spurious properties in interaction networks and can affect their clustering coefficient and average shortest path length. 
Here we investigate whether the assortativity coefficient $a$ (Eq.~\ref{eq:assort}) is also affected by these influences. 
We make use of a model which implements the null hypothesis that time series are obtained from independent stochastic processes\cite{Bialonski2011b}. 

Let $z_i$ be time series whose entries $z_i(t)$,$i \in \{1,\ldots, N\}$, are independently drawn from a uniform probability distribution $\mathcal{U}$ on the interval $(0,1)$. 
By choosing different length $T$ of time series and deriving interaction networks from $z_i$ as described in the methods section, we can study the influence of the length of time series on the assortativity coefficient. 
We additionally introduce the possibility to obtain time series $x_i$ with varying frequency content (i.e., varying serial correlations) by applying a moving average, 
\begin{equation}\label{eq:model}
 x_{i,M,T}(t) = M^{-1} \sum_{l=t}^{t+M-1} z_i(l),\qquad z_i(l) \sim \mathcal{U}\text{,}
\end{equation}
where $M$ is the size of the moving average and $T$ is the length of time series. 
For $M=1$, no serial correlations are introduced and we obtain $x_{i,1,T}(t) = z_i(t)\forall t$. For $M > 1$, the moving average acts as a low-pass filter. By choosing different values of $M$ ($M\ll T$) and keeping $T=T^\prime$ constant, we can study the influence of the frequency content of time series on the assortativity coefficient. 
We note that $x_{i,M,T}$ and $x_{j,M,T}$ are independent for $i\neq j$ by construction. 

We generate an ensemble of $E=1000$ networks with $N=100$ nodes each from the time series as defined in equation~\eqref{eq:model} for given values of $M$ and $T$. 
Each realization $e\in\{1,\ldots,E\}$ of a network is derived by thresholding the values of signal interdependence \rhoc{ij} or \rhom{ij} ($i,j \in \{1,\ldots,N\}$) obtained from the $e$-th realization of pairs of time series ($x^{(e)}_{i,M,T}$, $x^{(e)}_{j,M,T}$). 
The threshold was determined such that the resulting network possesses a link density $\epsilon = 0.1$. 
With $a_c^{(e)}(M,T)$ and $a_m^{(e)}(M,T)$ we denote the assortativity coefficient determined from the $e$-th realization of a network which was derived based on \rhoc{} or \rhom{}, respectively.
Mean values determined from all $E$ realizations are denoted as \hac{M,T} and \ham{M,T}, respectively.

In Fig.~\ref{fig:3} we show the dependence of \hac{M,500} (left panel) and \ham{M,500} (right panel) on the size $M$ of the moving average for time series of length $T^\prime=500$. 
Remarkably, for increasing values of $M$ and thus for an increasing relative amount of low-frequency contributions in time series, we observe the assortativity coefficient of interaction networks to increase. 
Likewise, if we keep $M=1$ constant and decrease the length of time series $T=T_\text{eff}$, we observe a similar dependency for \hac{1,T_\text{eff}} and \ham{1,T_\text{eff}} on $T$: the assortativity coefficient takes on larger values the shorter the time series. 
We note that the similarity between \hac{M,500} and \hac{1,T_\text{eff}} may be traced back to variances of the correlation coefficients  between time series $x^{(r)}_{i,1,T}$ and $x^{(r)}_{j,1,T}$ as well as between $x^{(r)}_{i,M,T^\prime}$ and $x^{(r)}_{j,M,T^\prime}$ which become approximately equal for $T=T_\text{eff}$, the effective length of time series,
\begin{equation*}
 T_\mathrm{eff}(M) = T^\prime \left(\frac{2}{3}M+\frac{1}{3M}\right)^{-1},
\end{equation*}
as argued in a previous study\cite{Bialonski2011b}.

To summarize, the finite length of time series and low-frequency contributions likely lead to interaction networks which would be classified as assortative networks according to the positiveness of the assortativity coefficient. 
This finding is particularly interesting in the light of network null models used to assess the significance of findings in field studies. 
The expectation value of the assortativity coefficient for frequently used null models like \ER networks\cite{Gilbert1959,Erdos1959,Erdos1960,Erdos1961} or random networks derived from the configuration model \cite{Bender1978,MolloyReed1995,Wormald1999} or by random link-switching \cite{Rao1996,Roberts2000,Maslov2002,Maslov2004,Randrup2005,Blitzstein2010,DelGenio2010} is zero (apart from effects due to a finite number of nodes). 
These models do not take into account the way how interaction networks are usually derived from multivariate field data. 
Although the question of which null model to choose is still a matter of an ongoing scientific debate\cite{Kramer2009,Bialonski2011b,Palus2011,Hlinka2012,Zalesky2012,Simpson2012,Fortunito2013}, we here use a null model\cite{Bialonski2011b} which accounts for the above demonstrated effects of finite length and low-frequency content of time series: networks are derived from surrogate time series. These surrogate time series possess the same amplitude distribution, the same number of sampling points and approximately the same frequency content as the empirical time series but are random in all other aspects\cite{Schreiber2000a}. 

\section{Derivation of equation (6)}

In the following we derive equation~\eqref{eq:assort} from the definition of the assortativity coefficient. Let us consider an undirected unweighted network consisting of $N$ nodes which is defined by its adjacency matrix $A$. To simplify the following steps without changing any of the results, we consider each undirected link of the network to be represented by two directed links. All directed links are contained in the set $E$, and the number of links is denoted as $|E|$. Each link $e\in E$ connects two nodes which possess degrees $l_e$ and $m_e$, respectively. 

The assortativity coefficient $a$ is defined as the Pearson correlation coefficient between the degrees of nodes at both ends of a link \cite{Newman2002a,Newman2003b}, i.e.
\begin{equation}\label{app:assort:eq}
 a = \frac{\mbox{Cov}(l,m)} {\sigma(l) \sigma(m)} = \frac{\overline{lm}-\overline{l}\,\overline{m}}{\sqrt{\overline{l^2}-\overline{l}^2}\sqrt{\overline{m^2}-\overline{m}^2}}\mbox{,}
\end{equation}
where Cov and $\sigma$ denote the covariance and the standard deviation, respectively. Note that $\overline{m} = \overline{l}$ and $\overline{m^2} = \overline{l^2}$ since our network is undirected. The mean values $\overline{l}$ and $\overline{l^2}$ are determined from sums over the links $e\in E$ which we can translate into sums over the nodes $i\in\{1,\ldots,N\}$ of the network by making use of the following observations: the number $|E|$ of directed links is equal to the sum of all degrees of nodes,
\begin{equation}\label{app:rel1}
 |E| = \sum_{i=1}^{N} k_i = K_1.
\end{equation}
The degree $k_i$ of a node $i$ is defined as $k_i = \sum_{j=1}^{N} A_{ij}$. Thus we obtain for the mean value $\overline{l}$ of the degrees of nodes at one end of links,
\begin{eqnarray}\label{app:rel2}
 \overline{l} &=& \frac{1}{|E|}\sum_{e=1}^{|E|} l_e = \frac{1}{K_1}\sum_{i=1}^{N}\sum_{j=1}^{N}A_{ij}k_i \nonumber\\ &=&\frac{1}{K_1}\sum_{i=1}^N k_i^2 = \frac{K_2}{K_1}\mbox{.}
\end{eqnarray}
Following the same line of reasoning, we obtain
\begin{eqnarray}\label{app:rel3}
 \overline{l^2} &=& \frac{1}{|E|}\sum_{e=1}^{|E|} l_e^2 = \frac{1}{K_1}\sum_{i=1}^{N}\sum_{j=1}^{N}A_{ij}k_i^2 \nonumber\\ &=&\frac{1}{K_1}\sum_{i=1}^N k_i^3 = \frac{K_3}{K_1}\mbox{.}
\end{eqnarray}
The mean value of the products of the degrees of nodes at both ends of links can be reformulated as
\begin{eqnarray}\label{app:mixedterm}
 \overline{lm} &=& \frac{1}{|E|}\sum_{e=1}^{|E|} l_e m_e = \frac{1}{K_1} \sum_{i=1}^{N} \sum_{j=1}^{N} A_{ij}k_i k_j\nonumber\\
 &=& \frac{2}{K_1} \sum_{i=1}^{N} \sum_{j=1}^{i-1} A_{ij}k_i k_j\mbox{,}
\end{eqnarray}
where the factor $2$ accounts for the fact that we considered directed links. Inserting equations~\eqref{app:rel2}, \eqref{app:rel3}, and \eqref{app:mixedterm} into equation~\eqref{app:assort:eq}, we finally obtain
\begin{equation}
 a = \left(K_1 K_3 - K_2^2 \right)^{-1} \left(2K_1 \sum_{i=1}^N \sum_{j=1}^{i-1} A_{ij}k_i k_j -K_2^2 \right).
\end{equation}

\end{document}